\documentstyle[aps,preprint,epsf]{revtex}

\makeatletter
\def\simleq{\mathrel{\mathpalette\gl@align<}}
\def\simgeq{\mathrel{\mathpalette\gl@align>}}
\def\gl@align#1#2{\lower.6ex\vbox{\baselineskip\z@skip\lineskip\z@
     \ialign{$\m@th#1\hfill##\hfil$\crcr#2\crcr\sim\crcr}}}
\makeatother

\newcommand{\fslash}[1]{\ooalign{\hfil/\hfil\crcr$#1$}}
\newcommand{\bra}{\langle}
\newcommand{\ket}{\rangle}
\newcommand{\braket}[1]{\bra #1 \ket}
\newcommand{\qq}{\braket{\bar{q}q}}
\renewcommand{\ss}{\braket{\bar{s}s}}
\newcommand{\GG}%
{ \left \langle {\alpha_s \over \pi}{\cal G}^2 \right \rangle }

\begin{document}
\draft
\tighten
\preprint{\vbox{\hfill TIT/HEP-447/NP}}

\title{Pertinent Dirac structure for QCD sum rules of 
 meson-baryon coupling constants}
\author{Takumi Doi$^1$ 
	\footnote{E-mail : doi@th.phys.titech.ac.jp},
Hungchong Kim$^2$ and 
Makoto Oka$^1$}
\address{$^1$ Department of Physics, Tokyo Institute of Technology, 
Tokyo 152-8551, Japan }
\address{$^2$ Institute of Physics and Applied Physics, Yonsei University, 
Seoul 120-749, Korea }

\maketitle


\begin{abstract}
Using general baryon interpolating fields $J_B$ for $B= N, \Xi, \Sigma, $ 
without derivative, we study QCD sum rules for meson-baryon couplings
and their dependence on Dirac structures for the two-point correlation 
function with a meson
$i\int d^4x e^{iqx} \bra 0|{\rm T}[J_B(x)\bar{J}_B(0)] |{\cal M}(p)\ket$.
Three distinct Dirac structures are compared:
$i\gamma_5$, $i\gamma_5\fslash{p}$, and
$\gamma_5\sigma_{\mu\nu}q^\mu p^\nu$ structures.  From the 
dependence of the OPE on general baryon interpolating fields,
we propose criteria for choosing an appropriate Dirac structure 
for the coupling sum rules.
The $\gamma_5\sigma_{\mu\nu}q^\mu p^\nu$ sum rules
satisfy the criteria while the $i\gamma_5$ sum rules
beyond the chiral limit do not.  For 
the $i\gamma_5\fslash{p}$ sum rules, 
the large continuum contributions prohibit reliable prediction
for the couplings. 
Thus, the $\gamma_5\sigma_{\mu\nu}q^\mu p^\nu$ structure
seems pertinent for realistic predictions.
In the SU(3) limit, we identify the OPE terms responsible for
the $F/D$ ratio. We then study the dependence of the
ratio on the baryon interpolating fields.
We conclude the ratio $F/D \sim 0.6-0.8$ for appropriate choice
of the interpolating fields.
\end{abstract}




\section{INTRODUCTION}
\label{sec:intro}

In QCD sum rule approaches~\cite{SVZ},
the two-point correlation function with a pion
\begin{eqnarray}
i\int d^4x e^{iq\cdot x} \bra 0|{\rm T}[J_N(x)\bar{J}_N(0)] |\pi (p)\ket
\end{eqnarray}
is often used to calculate the $\pi NN$ 
coupling~\cite{SH,BK,hung1,hung2,hung5,KD1,KD2} by facilitating a
general external field method developed in Ref.~\cite{ioffe-smilga}. 
This correlation function contains three distinct Dirac structures
(1) $i\gamma _5$ (PS), (2) $\gamma_5\sigma_{\mu\nu}q^\mu p^\nu$ (T), 
and (3) $i\gamma _5\fslash{p}$ (PV), each of which can
in principle be used to calculate the coupling.
Currently, there is an issue of the Dirac structure
dependence of the sum rule results~\cite{hung1,hung2}.
In calculating the coupling, one can
construct either
the PS sum rules beyond the chiral limit~\cite{hung5,KD1}
or the T sum rules~\cite{hung1,KD2}.
Both sum rules yield the $\pi NN$ coupling close to
its empirical value.
On the other hand, the $i\gamma _5\fslash{p}$ sum rules
contain large contributions from the continuum, which
therefore do not provide reliable results. 

The PS and T sum rules have been extended to calculate the
meson-baryon couplings $\eta NN$, $\pi \Xi\Xi$, $\eta \Xi\Xi$, 
$\pi\Sigma \Sigma$ and $\eta\Sigma\Sigma$~\cite{KD1,KD2} by
considering the two-point correlation function with
a meson,
\begin{eqnarray}
i\int d^4x~ e^{iq\cdot x} 
\bra 0|{\rm T}[J_B(x)\bar{J}_B(0)] |{\cal M}(p)\ket\ .
\label{mcor}
\end{eqnarray}
Calculation of the couplings from this correlation function
is somewhat limited due to the ignorance of meson wave functions
when heavier mesons are involved.  In the SU(3) limit however,
this correlation function can be used to determine 
the so-called $F/D$ 
ratio unambiguously because in this limit the OPE can be exactly 
classified~\cite{KD1,KD2} according to SU(3) relations for the
couplings~\cite{swart}.  The $F/D$ ratio is an important input
in making realistic potential models for hyperon-baryon 
interactions~\cite{rijken1,rijken2} as well as in analyzing
the hyperon semileptonic data.

At present, there is a clear Dirac structure dependence
in the calculation of the $F/D$ ratio using Eq.~(\ref{mcor}).   
In particular, we have reported from the PS sum rules  
$F/D\sim 0.2$~\cite{KD1}
while from T sum rules $F/D\sim 0.78$~\cite{KD2}.
Thus, even though the two sum rules with different Dirac
structures were successful in reproducing the empirical
$\pi NN$ coupling, their prediction for the $F/D$
ratio is quite different.

To resolve this issue, additional criteria to choose
a proper Dirac structure are needed for reliable
predictions on the $F/D$ ratio as well as the
meson-baryon couplings.
For this purpose, we first note that in Ref.~\cite{KD1,KD2}
the Ioffe current or its
SU(3) rotated version has been used to
construct sum rules Eq.~(\ref{mcor}).
The Ioffe current however is a specific choice for 
the nucleon current among infinitely many possibilities.
The Ioffe current is often used for the nucleon
because it gets large contributions from the 
chiral breaking parameter $\qq$. 
In addition, direct instantons are believed to play less roles
in this current.

Nevertheless, it may be useful to study 
the dependence of the sum rule results
on general baryon currents. 
Depending on the currents,
it is expected that the overlap $\lambda_B$ between the physical baryon
state and the current may be altered but ideally the
physical parameters such as meson-baryon couplings remain unchanged.
Indeed, from the correlation function Eq.~(\ref{mcor}),
what will actually be determined is the overlap strength multiplied by
the coupling of concern. 
In the  SU(3) symmetric limit, all the strengths
depend only on the currents.
They are determined from the corresponding baryon mass sum rules
and all the baryon masses are the same in the
SU(3) limit. Thus, in this limit,
the dependence on the currents should be driven
by the common overlap strength, which in return provides the
coupling independent of the currents. 
This ideal aspect will be pursued in this work as
a criterion for choosing a proper Dirac structure.

An alternative way is to calculate baryon axial charges 
and convert them into meson-baryon couplings using the 
Goldberger-Treiman relation.
Ref.~\cite{bely-kogan} considered the nucleon correlation
function in external axial vector field and constructed 
a sum rule for $g_A -1$ using one specific Dirac structure.
Recently, a new approach was proposed in Ref.~\cite{nishikawa-kondo}
where the axial vector correlation function in a one-nucleon state
is considered.
Both obtained an excellent agreement for $g_A$ of the nucleon.

This paper is organized as follows.
In Section~\ref{sec:coupling-qsr}, we construct meson-baryon
coupling sum rules using general baryon currents. 
A brief discussion on the OPE based on chirality is given  in
Section~\ref{sec:chirality}.
We then briefly check in Section~\ref{sec:s0-dep} whether
the discussion on the continuum threshold~\cite{hung2,KD2}
is still valid when the general baryon currents are used in the sum 
rules.
In Section~\ref{sec:current-dep}, 
the dependence of the OPE on the baryon currents is studied.
We study in the SU(3) limit whether or not 
the dependence on the currents are mostly contained in
the overlap $\lambda_B$.
This constraint gives us a new criterion
to choose an appropriate Dirac structure.
In Section~\ref{sec:su3}, we calculate the couplings 
in the SU(3) limit from the $\gamma_5\sigma_{\mu\nu}q^\mu p^\nu$ 
structure.
The $F/D$ ratio is identified in terms of the OPE.
Conclusions are given in Section~\ref{sec:summary}.



\section{CONSTRUCTION of the  QCD SUM RULES}
\label{sec:coupling-qsr}

We use the two-point correlation function with a meson,
\begin{eqnarray}
i\int d^4x e^{iq\cdot x} \bra 0|{\rm T}[J_B(x)\bar{J}_B(0)] 
|{\cal M}(p)\ket
\label{tmcor}
\end{eqnarray}
where $J_B$ is the baryon current of concern and $p$ is the momentum of 
meson ${\cal M}$. Meson states $\pi$ and $\eta$, and baryon currents for
the proton, $\Xi$ and $\Sigma$ will be considered in this work.  

The proton current is constructed from two $u$-quarks and one 
$d$-quark by assuming that all three quarks are in the s-wave state.
In the construction of the current, one up and one down quark are  
combined into an
isoscalar diquark. The other up quark is attached to the
diquark so that quantum numbers of the proton are
carried by the attached up quark. In this method, there
are two possible combinations for the current.  
The general proton current is a linear combination of
the two possibilities mediated by a real parameter $t$,
\begin{eqnarray}
J_p(x;t) &=& 2\epsilon _{abc}
[\ (u_a^T(x) C d_b(x))\gamma_5 u_c(x) 
+ t\ (u_a^T(x) C\gamma_5 d_b(x))u_c(x)\ ]\ . 
\label{eq:proton_current}
\end{eqnarray}
Here, $a,b,c$ are color indices, $T$ denotes the transpose 
with respect to the Dirac indices, and $C$ the charge conjugation.
The choice $t=-1$ is called the Ioffe current~\cite{ioffe}.
The currents for $\Xi$ and $\Sigma$ are obtained from the
proton current via SU(3) rotations~\cite{qsr},
\begin{eqnarray}
J_\Xi(x;t) &=& -2\epsilon _{abc}
[\ (s_a^T(x) C u_b(x))\gamma_5 s_c(x) 
+ t\ (s_a^T(x) C\gamma_5 u_b(x))s_c(x)\ ]\ , \nonumber\\
J_\Sigma(x;t) &=& \ \ 2\epsilon _{abc}
[\ (u_a^T(x) C s_b(x))\gamma_5 u_c(x) 
+ t\ (u_a^T(x) C\gamma_5 s_b(x))u_c(x)\ ]\ . 
\label{eq:current}
\end{eqnarray}

When going beyond the soft-meson limit, 
one can consider three distinct Dirac structures in 
correlation function in constructing sum rules: 
$i\gamma_5$ (PS), $\gamma_5\sigma_{\mu\nu}q^\mu p^\nu$ (T) and 
$i\gamma_5\fslash{p}$ (PV).
For the $i\gamma_5$ structure, the sum rules are
constructed at the order $p^2 = m_\pi^2\ $~\cite{hung5}. 
At this order, the terms linear in quark mass $(m_q)$ in
the OPE should be included because $m_q$ is the same
chiral order with $m_\pi^2$ via 
the Gell-Mann--Oakes--Renner relation,
\begin{eqnarray}
-2m_q\qq = m_\pi^2 f_\pi^2 \ .
\end{eqnarray}
On the other hand, for the T and PV structures, 
we construct the sum rules at the order ${\cal O}(p)$.
At this order,
the $m_q$ terms should not be included in the OPE.
Technical details on the OPE calculation can be found in 
Refs.~\cite{KD1,KD2}.

In constructing the phenomenological side,
we first define $\lambda_B(t)$,
the coupling strength between the baryon current $J_B(x;t)$ and 
the physical baryon field $\psi_B(x)$.
Using the pseudoscalar type interaction between the meson
and baryons $g_{{\cal M}B}\bar{\psi}_B i\gamma_5 \psi_B {\cal M}$,
we obtain the phenomenological side of the correlation function:
\begin{eqnarray}
&i\gamma_5\ \mbox{structure}&\ \mbox{at the order}\ {\cal O}(p^2)
\qquad 
i\gamma_5
p^2 \frac{g_{{\cal M}B}\lambda_B^2(t)}{(q^2-m_B^2)^2} + \cdots \ , \\
&\gamma_5\sigma_{\mu\nu}q^\mu p^\nu\ \mbox{structure}&\ \mbox{at the order}\ 
{\cal O}(p)
\qquad 
\gamma_5\sigma_{\mu\nu}q^\mu p^\nu 
\frac{g_{{\cal M}B}\lambda_B^2(t)}{(q^2-m_B^2)^2} + \cdots \ , \\
&i\gamma_5\fslash{p}\ \mbox{structure}&\ \mbox{at the order}\ {\cal O}(p)
\qquad 
-i\gamma_5\fslash{p} 
\frac{g_{{\cal M}B}\lambda_B^2(t)m_B}{(q^2-m_B^2)^2} + \cdots \ .
\end{eqnarray}
The ellipsis denotes contributions from higher resonances as
well as a single pole associated with transitions from
the ground state to higher resonances.
The continuum contributions come from transitions
among higher resonances, whose spectral densities
are modeled with a step function starting at the threshold $S_0$. 
Matching the OPE side with the phenomenological side and taking 
Borel transformation~\footnote{Note, we use a single dispersion
relation as advocated in Ref.~\cite{com,single}.}, 
we get the sum rules of the form 
\begin{eqnarray}
g_{{\cal M}B}\lambda_B^2(t)\left[\ 1+ A_{{\cal M}B}(t) M^2 \ \right] 
&=& 
e^{m_B^2/M^2}\cdot
F^{\mbox{\tiny OPE}}_{\mbox{\tiny\it ${\cal M}$B}}(M^2;t) 
\quad \equiv \quad
f^{\mbox{\tiny OPE}}_{\mbox{\tiny\it ${\cal M}$B}}(M^2;t) 
\label{eq:qsr-formula}
\end{eqnarray}
where the single pole term in the phenomenological side has been denoted 
by $A_{{\cal M}B}$. 
Expressions for the OPE 
$F^{\mbox{\tiny OPE}}_{\mbox{\tiny\it ${\cal M}$B}} (M^2;t)$ 
are given in the
Appendix~\ref{app:coupling-qsr}.



\section{Chirality consideration}
\label{sec:chirality}

The OPEs given in the Appendix~\ref{app:coupling-qsr} have an interesting feature
to discuss when $t=1$.   Specifically, in the $i\gamma_5$ and
$\gamma_5 \sigma_{\mu\nu} q^\mu p^\mu$ sum rules, Wilson coefficients
of chiral-odd operators $\qq$, $f_{3\pi}$, $\qq \GG$, and $m_0^2 \qq$ are all
zero when $t=1$.  Also in the $i\gamma_5\fslash{p}$ sum rules,
contributions from
the chiral-even operators $\qq^2$ and $m_0^2 \qq^2$ are zero.  
To understand this feature, it is useful to decompose the
correlator according to chirality of the current,
\begin{eqnarray}
J_B\bar{J}_B = J^R_B \bar{J}^R_B + J^R_B \bar{J}^L_B + 
J^L_B \bar{J}^R_B + J^L_B \bar{J}^L_B \ .
\end{eqnarray}
$J^L_B (J^R_B)$ denotes the left-handed (right-handed) component of the
current $J_B$.   On the other hand, Eq.~(\ref{tmcor}) can be written 
\begin{eqnarray}
i\int d^4x e^{iq\cdot x} \bra 0|{\rm T}[J_B(x)\bar{J}_B(0)] 
|{\cal M}(p)\ket &=& i\gamma_5~\Pi_{\rm ps} 
+ i\gamma_5\fslash{p}~\Pi_{\rm pv} 
\nonumber \\
&+& \gamma_5 \sigma_{\mu\nu} q^\mu p^\nu~\Pi_{\rm T} \ .
\end{eqnarray}
Thus, it is easy to see that
the $i\gamma_5$ and $\gamma_5 \sigma_{\mu\nu}$ structures have 
nonzero contributions only from the chiral mixing term 
$J^R \bar{J}^L + J^L \bar{J}^R$, while the chirality conserving term
$J^R \bar{J}^R+J^L \bar{J}^L$ contributes only to
the $i\gamma_5\fslash{p}$ structure.

Now let us classify QCD operators contributing to each Dirac structure. 
To do that, we suppress for simplicity the color indices 
and write baryon current as 
\begin{equation}
J \sim (q^T C q)\gamma_5 q + t\ (q^T C \gamma_5 q)q\ .
\end{equation}
Here $q=u,d,s$.
When $t=1$, it is straightforward to show that
\begin{eqnarray}
J_R &\sim&  2(q_R^T C q_R )q_R \label{cr}\ , \\
J_L &\sim& -2(q_L^T C q_L )q_L \label{cl}\ .
\end{eqnarray}
Thus, at this specific $t$, chirality of all quarks are the same as 
that of the baryon.

In the $i\gamma_5$ or $\gamma_5 \sigma_{\mu\nu}$ sum rules,
we need to consider the products $J^R \bar{J}^L$ and $J^L \bar{J}^R$.  
In making such products
using Eqs.~(\ref{cr}) (\ref{cl}), all three quark propagators
should break the chirality when they move from the coordinate $0$ to $x$.
Hence, it is easy to see that, among chiral-odd operators,
terms such as
\begin{equation}
m_q \qq ^2 , \qq ^3 , m_q^2 \qq , \cdots
\end{equation}
can contribute to the $i\gamma_5$ or $\gamma_5 \sigma_{\mu\nu}$ correlator, 
while other chiral-odd operators 
such as $\qq$, $f_{3\pi}$, $\qq \GG$, $m_0^2 \qq$ cannot.
 
On the other hand, in the $i\gamma_5 \fslash{p}$ structure,
the product $J^L \bar{J}^L$ or $J^R \bar{J}^R$ contributes to
the sum rule.  
Among chiral-even operators, an operator
such as $\qq^2$ cannot be formed in the product 
$J^L \bar{J}^L$ or $J^R \bar{J}^R$  simply because 
two quarks with the same chirality cannot be combined
into the quark-antiquark pair.  Similarly, $m_0^2 \qq^2$ can not
be formed. This  explains the disappearance of such
terms in the OPE when $t=1$.



\section{Criterion I :
sensitivity to the continuum threshold}
\label{sec:s0-dep}

We now analyze sum rules of the three different Dirac structures
with the general baryon currents, Eqs~(\ref{eq:proton_current}) and
(\ref{eq:current}).
As pointed out in Refs.~\cite{hung2,KD2},
sum rule results from the $i\gamma_5\fslash{p}$ structure
are sensitive to the continuum threshold $S_0$
and therefore this structure is not reliable.
On the other hand, $i\gamma_5$ and 
$\gamma_5\sigma_{\mu\nu}q^\mu p^\nu$ structures
are insensitive to $S_0$.  
The chirality consideration suggested in
Ref.~\cite{hung2} implies that in the $i\gamma_5\fslash{p}$
sum rules the large slope and
the strong sensitivity to $S_0$ of the Borel curves
can be explained if higher resonances with different parities
add up.  With this scenario,
the higher resonances contributions cancel
each other in the  $i\gamma_5$ and $\gamma_5\sigma_{\mu\nu}q^\mu p^\nu$
sum rules therefore explaining the weak sensitivity to $S_0$ and
the small slope of the Borel curves.

\begin{figure}[hbtp]
 \caption{The Borel curve for the $\pi NN$ coupling from the $i\gamma_5$
   structure.
   $g_{\pi N}\lambda_N^2(t)$ 
   is determined by taking the intersection
   of the vertical axis ($M^2=0$) with the best-fitting linear curve.
   (See also Section~\ref{sec:current-dep}).
   The thick lines are for $S_0 = 2.07\ {\rm GeV}^2$ 
   while the thin lines are for $S_0 = 2.57\ {\rm GeV}^2\ $. 
   The three different sets of curves correspond to three different
   values of $t$. The long-dashed lines are for $t=1.5$, 
   the solid lines for $t=-1.5$ and the dot-dashed lines for $t=-1.0$.
   The difference by changing the continuum threshold is 
   only $2-3\%$ at $M^2 = 1\ {\rm GeV}^2$ for each $t$.
}

  \epsfysize = 85mm
  \centerline{\epsfbox{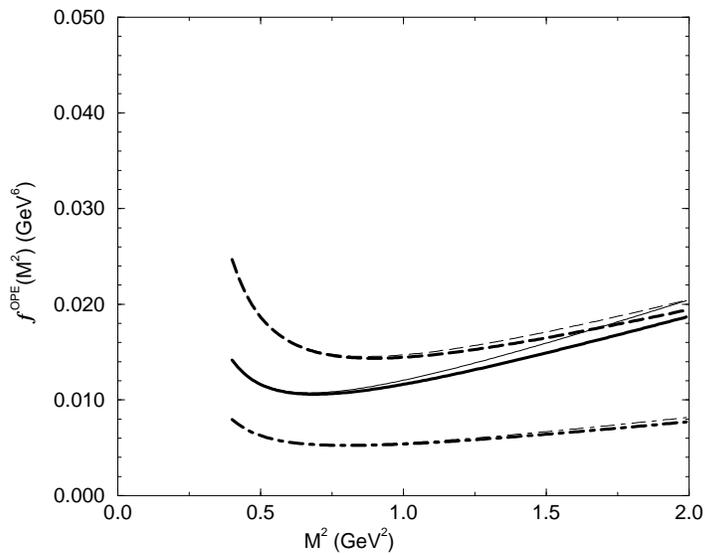}}

 \label{fig:s0_dep-PS}
\eject
\end{figure}

\begin{figure}[hbtp]
 \caption{The Borel curve for the $\pi NN$ coupling from the
   $\gamma_5\sigma_{\mu\nu}q^\mu p^\nu$ structure.
   The thick lines are for $S_0 = 2.07\ {\rm GeV}^2\ $ case,
   while thin lines are for $S_0 = 2.57\ {\rm GeV}^2\ $case.
   The long-dashed, solid, or dot-dashed lines
   correspond to  $t = 1.5,-1.5,-1.0$, respectively.
   The difference by changing the continuum threshold is 
   only $2-3\%$ level at $M^2 = 1\ {\rm GeV}^2$ for each $t$.
}
  \epsfysize=85mm
  \centerline{\epsfbox{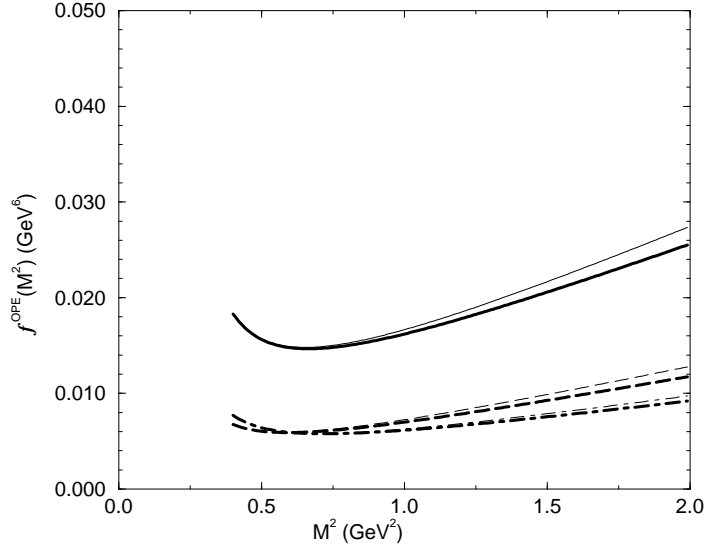}}
  
 \label{fig:s0_dep-T}
\end{figure}

\begin{figure}[hbtp]
 \caption{The Borel curve for the $\pi NN$ coupling from the
   $i\gamma_5 /\!\!\!p$ structure. Each curve is obtained similarly
   as the $i\gamma_5$ and $\gamma_5\sigma_{\mu\nu}q^\mu p^\nu$ cases. 
   The difference by changing the continuum threshold is 
   large, almost $15\%$ level at $M^2 = 1\ {\rm GeV}^2$.
}

  \epsfysize=85mm
  \centerline{\epsfbox{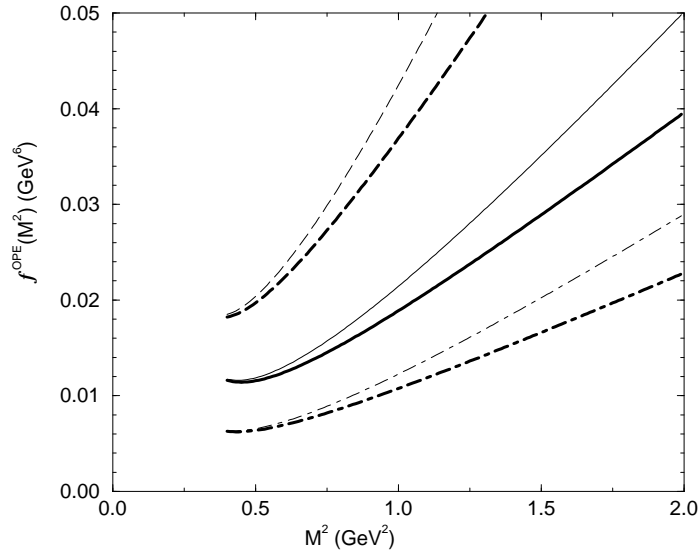}}

 \label{fig:s0_dep-PV}
\eject
\end{figure}

Since only the Ioffe current is used in the analysis of Refs.~\cite{hung2,KD2},
let us briefly check if this scenario still works when the
general baryon currents are used.  
As the scenario does not
rely on the specific form for the current, what has been claimed in
Refs.~\cite{hung2,KD2} must be valid even with
the general baryon currents.
To see this, we plot the RHS of Eq.~(\ref{eq:qsr-formula}) for the $\pi NN$ 
coupling from $i\gamma_5$, $\gamma_5\sigma_{\mu\nu}q^\mu p^\nu$ and
$i\gamma_5\fslash{p}$ structures
in Figs.~\ref{fig:s0_dep-PS},~\ref{fig:s0_dep-T},~\ref{fig:s0_dep-PV},
respectively. To show the dependence on $t$, we
plot the curves for $t= -1.5, 1.5$ as well as $t=-1.0$ (the Ioffe current).
In these plots, we use the standard QCD parameters,
\begin{eqnarray}
\qq &=& -(0.23~ {\rm GeV})^3\;; \quad
\GG
= (0.33~{\rm GeV})^4 \ ,
\nonumber \\
\delta^2&=&0.2~{\rm GeV}^2\;; \quad m_0^2=0.8~{\rm GeV}^2\ .
\label{eq:qcd-parameter}
\end{eqnarray}
For each $t$, 
the thick lines are for the continuum threshold
$S_0 = 2.07\ {\rm GeV}^2$
corresponding to the Roper resonance,
while the thin lines for $S_0 = 2.57\ {\rm GeV}^2$.
The trend observed here is the same for the other couplings.

In Fig.~\ref{fig:s0_dep-PV},
we observe that the $i\gamma_5\fslash{p}$ structure is  
sensitive to the continuum threshold even when
the general current is used.
The difference by changing the continuum threshold
is $\sim 15\%$ at $M^2 = 1~{\rm GeV}^2$.  Note also that
the slope is relatively large in this case.
Since the coupling is determined from the intersection of the
best fitting curve with the vertical line at $M^2=0$
(see Eq.~(\ref{eq:qsr-formula})), the $15 \%$ change
at  $M^2 = 1~{\rm GeV}^2$, when it combined with the large slope,
produces huge change in the extracted coupling.    
In contrast, from Figs.~\ref{fig:s0_dep-PS} and ~\ref{fig:s0_dep-T},
the $i\gamma_5$ and $\gamma_5\sigma_{\mu\nu}q^\mu p^\nu$ structures
are insensitive to $S_0$. Also the
slopes of the curves are small.
This observation is practically independent of the parameter $t$.
At $M^2 = 1~{\rm GeV}^2$, the difference is 
only $2-3\%$ level.
Thus, the analysis in Refs.~\cite{hung2,KD2} is still valid and
the sum rule results from $i\gamma_5\fslash{p}$ 
structure should be discarded under this consideration.



\section{Criterion II : the dependence of the OPE on baryon currents}
\label{sec:current-dep}

Using the sum rules derived in 
Section~\ref{sec:coupling-qsr},
we discuss the dependence of the OPE 
on the baryon current (i.e. the dependence on $t$).
For a given $t$, we linearly fit
the RHS of 
Eq.~(\ref{eq:qsr-formula})
\begin{eqnarray*}
g_{{\cal M}B}\lambda_B^2(t)\left[\ 1+ A_{{\cal M}B} (t) M^2 \ \right] 
\ = \ 
f^{\mbox{\tiny OPE}}_{\mbox{\tiny\it ${\cal M}$B}}(M^2;t)\ ,
\end{eqnarray*}
and determine 
$\left[ g_{{\cal M}B}\lambda_B^2(t) \right]_{\mbox{\scriptsize fitted}}$.
Because $f^{\mbox{\tiny OPE}}_{\mbox{\tiny\it ${\cal M}$B}}$ is 
quadratic in $t$, 
$\left[ g_{{\cal M}B}\lambda_B^2(t) \right]_{\mbox{\scriptsize fitted}}$
is also quadratic. 
Ideally, the physical parameter $g_{{\cal M}B}$ should
be independent of $t$  if the sum rules are reliable.
In other words, $t$ is just a parameter for the current. By
changing $t$, only the coupling strength $\lambda_B^2(t)$
is expected to be affected, but not the physical parameter.
This is a constraint to be satisfied when the sum rules are ``good.''

To proceed, we take the SU(3) symmetric limit. Then,
the strength $\lambda_B(t)$ should be independent of the baryons,
\begin{eqnarray}
\lambda_N(t) = \lambda_\Xi (t) = \lambda_\Sigma (t)\ ,
\end{eqnarray}
as the baryon mass sum rules are the same in the limit.
Furthermore, we have
\begin{eqnarray}
&\ss = \qq &\;; \quad m_\eta = m_\pi \ , \nonumber\\
&f_\eta = f_\pi &\;; \quad f_{3\eta} = f_{3\pi} \ , \nonumber\\
&m_N = m_\Xi = m_\Sigma &\;; \quad m_s = m_q\ .
\label{eq:su3-parameter}
\end{eqnarray}
This SU(3) limit is particularly interesting when we select
a suitable Dirac structure. 
Suppose we plot 
$\left[ g_{{\cal M}B}\lambda_B^2(t) \right]_{\mbox{\scriptsize fitted}}$
in terms of $t$.  If the sum rules are ``good'',
all $g_{{\cal M}B}$ should be just constants, independent of $t$. 
The functional behavior is driven only by 
the strength $\lambda_B^2(t)$.
The baryon mass sum rules in the SU(3) limit constrain that
all $\lambda_B^2(t)$ are {\it the same} irrespective of the baryons. 
%
%
Therefore,
``good'' sum rules must give 
$\left[ g_{{\cal M}B}\lambda_B^2(t) \right]_{\mbox{\scriptsize fitted}}$
which are proportional to each other.

Our constraint should be satisfied when the OPE  are
exact.
But in practice, the full OPE terms $f^{\rm OPE}_{\rm full}$
are separated into two groups, 
\begin{eqnarray}
f^{\rm OPE}_{\rm full} = f^{\rm OPE}_{\rm calc} + f^{\rm OPE}_{\rm rest}\ ,
\end{eqnarray}
where $f^{\rm OPE}_{\rm calc}$
denotes the calculable OPE,
and $f^{\rm OPE}_{\rm rest}$
denotes the rest of the full OPE.
In this notation, the reliability of sum rule simply means 
\begin{eqnarray}
f^{\rm OPE}_{\rm calc} \gg f^{\rm OPE}_{\rm rest}\ .
\end{eqnarray}
The sum rules are ``unreliable'' if 
\begin{eqnarray}
f^{\rm OPE}_{\rm calc} \sim f^{\rm OPE}_{\rm rest}\ .
\end{eqnarray}
In the former case, we expect that
$\left[ g_{{\cal M}B}\lambda_B^2(t) \right]_{\mbox{\scriptsize fitted}}$
derived from $f^{\mbox{\tiny OPE}}%
_{\mbox{\tiny\it ${\cal M}$B;\rm\scriptsize\ calc}}$ 
is almost the same as those from 
$f^{\mbox{\tiny OPE}}_{\mbox{\tiny\it ${\cal M}$B;\rm\scriptsize\ full}}$
in most region of $t$.
On the other hand, in the latter case,
$\left[ g_{{\cal M}B}\lambda_B^2(t) \right]%
_{\mbox{\scriptsize fitted}}$
derived from $f^{\mbox{\tiny OPE}}%
_{\mbox{\tiny\it ${\cal M}$B;\rm\scriptsize\ calc}}$ 
may be quite different from those obtained from 
$f^{\mbox{\tiny OPE}}_{\mbox{\tiny\it ${\cal M}$B;\rm\scriptsize\ full}}$,
and our ideal constraint may not be satisfied in most $t$.

Therefore, the ideal constraint can be used as
a new criterion for choosing reliable sum rules.
In order to apply this constraint to our sum rules,
we again use the standard QCD parameters 
Eq.~(\ref{eq:qcd-parameter})
and linearly fit  
$\left[ g_{{\cal M}B}\lambda_B^2(t) \right]_{\mbox{\scriptsize fitted}}$
at each $t$.
In the fitting, the continuum threshold is set to 
$S_0 = 2.07\ {\rm GeV^2}$,  
corresponding to the Roper resonance, and 
the Borel window is taken 
$0.65 \leq M^2 \leq 1.24\ {\rm GeV^2}\ $
as in Refs.~\cite{hung5,KD1,KD2}.
In this Borel window,
(1) the Borel curve for each coupling is almost linear
(see Figs.~\ref{fig:s0_dep-PS},~\ref{fig:s0_dep-T}),
(2) the contribution from the highest dimensional OPE term is typically 
$5-15\%$ in the $\gamma_5\sigma_{\mu\nu}q^\mu p^\nu$ sum rules 
and $20\%$ level in the $i\gamma_5$ sum rules, and
(3) 
the continuum contribution is less than $20\%$
in both structures.
It should be noted that because all the couplings are related under
SU(3) rotations, 
we need to take a common Borel window~\cite{KD1,KD2}.

\begin{figure}[hbtp]
 \caption{
  $\left[ g_{{\cal M}B}\lambda_B^2(t) \right]_{\mbox{\scriptsize fitted}}$
  from the $i\gamma_5$ structure
  is plotted as a function of $t$,
  for $\pi NN$, $\eta NN$, $\pi\Xi\Xi$, $\eta\Xi\Xi$, 
  $\pi\Sigma\Sigma$, and $\eta\Sigma\Sigma$.
  We choose the Borel window as $0.65 \leq M^2 \leq 1.24\ {\rm GeV^2}\ $,
  and the continuum threshold as $S_0 = 2.07\ {\rm GeV^2}$.
}

  \epsfysize=85mm
  \centerline{\epsfbox{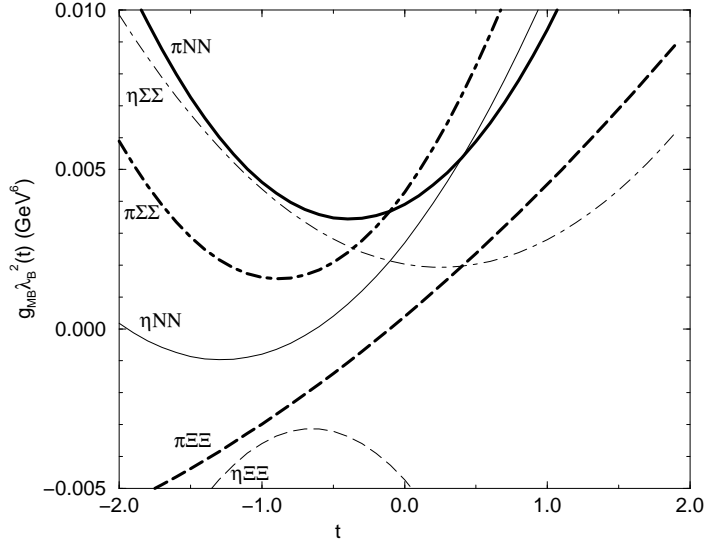}}

 \label{fig:t_dep-PS}
\end{figure}

\begin{figure}[hbtp]
 \caption{
  $\left[ g_{{\cal M}B}\lambda_B^2(t) \right]_{\mbox{\scriptsize fitted}}$
  from the $\gamma_5\sigma_{\mu\nu}q^\mu p^\nu$ structure
  is plotted as a function of $t$.
}
 \epsfysize=85mm
 \centerline{\epsfbox{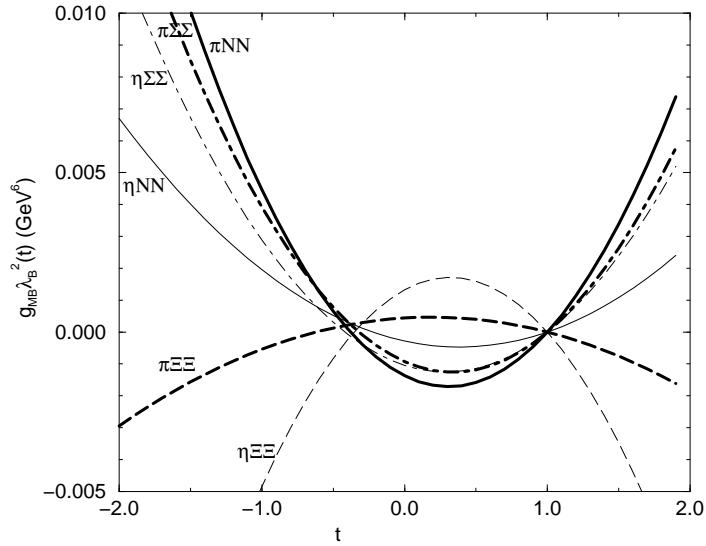}}

 \label{fig:t_dep-T}
\eject
\end{figure}

Fig.~\ref{fig:t_dep-PS} shows 
$\left[ g_{{\cal M}B}\lambda_B^2(t) \right]_{\mbox{\scriptsize fitted}}$
as a function of $t$
for the $i\gamma_5$ sum rules.
The $\gamma_5\sigma_{\mu\nu}q^\mu p^\nu$  cases
are shown in Fig.~\ref{fig:t_dep-T}.
Interesting features in the $\gamma_5\sigma_{\mu\nu}q^\mu p^\nu$  cases
are that (1) all the curves are zero when $t=1$
and almost zero at $t\sim -0.5$, (2) each extremum of the curves 
coincides around $t\sim 0.3$.
Under the chirality consideration given in Section ~\ref{sec:chirality},
we can easily understand why
$\left[ g_{{\cal M}B}\lambda_B^2(t) \right]_{\mbox{\scriptsize fitted}}$
is zero when $t=1$.  From the figure, though not exact, one 
observes that the curves can be almost overlapped
when multiplied by appropriate constants.
For example, let us compare the $\pi NN$ and $\eta NN$ curves.
When they are positive, the $\pi NN$ curve lies above
the $\eta NN$ curve.  When they are negative, the situation
is reversed. This behavior of the $\eta NN$ curve can be reproduced by
multiplying an  
appropriate constant to the $\pi NN$ curve. Of course,
this claim can not be made when $t\sim -0.5$ because
one curve becomes zero while the other does not.
Therefore, except around $t\sim -0.5$, 
the Borel curves satisfy the ideal constraint in most region
of $t$. 
Such a trend can not be observed from the
$i\gamma_5$ sum rules. (see Fig.~\ref{fig:t_dep-PS}.) 
Therefore, we claim that 
the $\gamma_5\sigma_{\mu\nu}q^\mu p^\nu$  sum rules
are more appropriate.

\begin{figure}[hbtp]
 \caption{$\lambda_B^2(t)$ is plotted with the thick solid line 
   as a function of $t$
   using chiral-odd nucleon mass sum rule 
   Eq.~(\ref{eq:mass-qsr}) at $M^2 = 1\ {\rm GeV^2}$ and
   the continuum threshold $S_0 = 2.07\ {\rm GeV^2}$.
   Also shown with the thin long-dashed line (the thin dot-dashed line)
   is for $M^2 = 1.2\ {\rm GeV^2},\ S_0 = 2.07\ {\rm GeV^2}$
   ($M^2 = 1\ {\rm GeV^2},\ S_0 = 2.57\ {\rm GeV^2}$). 
}
  \epsfysize=85mm
  \centerline{\epsfbox{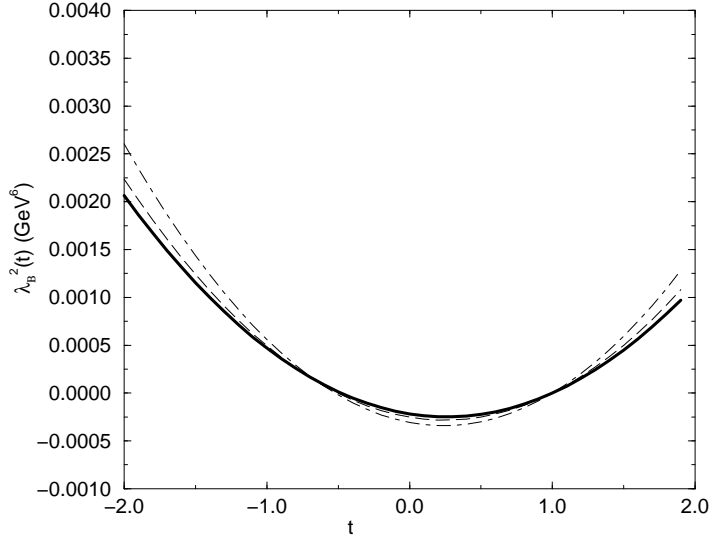}}

 \label{fig:t_dep-mass-qsr}
\end{figure}

To support our claim that the $\gamma_5\sigma_{\mu\nu}q^\mu p^\nu$ sum rules
are more suitable than those from the $i\gamma_5$ structure,
one more check to do is to see the $t$-dependence of $\lambda_B^2(t)$
from baryon mass sum rules.
In Fig.~\ref{fig:t_dep-mass-qsr},
$\lambda_B^2(t)$ in the SU(3) limit
is plotted using chiral-odd nucleon mass sum 
rule~\footnote{Definition of the function $E_n(x\equiv S_0/M^2)$ is given 
in the Appendix~\ref{app:coupling-qsr}.
The Wilson coefficient of the dimension 7 OPE  is different from
Ref.~\cite{leinweber}. When $t=-1$ (the Ioffe current), however,
our Wilson coefficient reduces to that of Ref.~\cite{ioffe-smilga}.
Nevertheless, the dimension 7 condensate contributes to the sum rule
only slightly. Thus, this discrepancy is marginal.}:
%
%
%
%
\begin{eqnarray}
m_B \lambda _B^2(t) e^{-m_B^2/M^2}
&=&
\frac{-4}{(2\pi )^4}
\left[\ \frac{\pi ^2}{4}(-5-2t+7t^2)\qq M^4 E_1(x) \right. 
\nonumber\\
&& 
+ \frac{3}{4}\pi^2 (1-t^2)m_0^2\qq M^2 E_0(x) \nonumber\\
&& \left.
+ \frac{\pi^4}{24}(-7+2t+5t^2)\qq\GG 
\ \right] \ ,
\label{eq:mass-qsr}
\end{eqnarray}
%
%
%
%
where $m_q$ order terms are neglected
and the SU(3) relations are used:
$m_N = m_\Xi = m_\Sigma \equiv m_B$; 
$\lambda_N = \lambda_\Xi = \lambda_\Sigma \equiv \lambda_B$;
$\braket{\bar{u}u} = \braket{\bar{d}d} = \ss \equiv \qq $.
Comparing with Fig.~\ref{fig:t_dep-T},
we confirm that the $t$-dependence of 
$\left[ g_{{\cal M}B}\lambda_B^2(t) \right]_{\mbox{\scriptsize fitted}}$
from the $\gamma_5\sigma_{\mu\nu}q^\mu p^\nu$ sum rules
can be reproduced from the $t$-dependence of $\lambda_B^2(t)$.

In the region $ -0.5 \simleq t \simleq 1$ in Fig.~\ref{fig:t_dep-mass-qsr}, 
$\lambda_B^2 (t)$ is negative,
thus not physical.
In this region, the sum rules should definitely fail  and a reliable
prediction for a physical parameter may not be possible.
At $t\sim -0.5$ or $t\sim 1$, of course, the OPE is almost zero
suggesting that there are cancellations among OPE terms,
i.e. the correlation function can not be well saturated
by the calculated OPE.
Therefore, the optimal current should be
chosen away from these points.
 


\section{The $F/D$ ratio from the pseudotensor sum rules}
\label{sec:su3}

In this section, we analyze the $\gamma_5 \sigma_{\mu\nu}q^\mu p^\nu$
sum rules to determine the $F/D$ ratio.
In particular, we investigate the $t$-dependence of the ratio using
the general interpolating fields for the baryons.
As already mentioned, mesons and baryons are classified according
to SU(3) symmetry, which provides simple relations for
the meson-baryon couplings in terms of the two parameters~\cite{swart} 
\begin{equation}
g_{\pi N}~~{\rm  and}~~ \alpha={F\over F+D}\ .
\end{equation}
That is,
\begin{eqnarray}
g_{\eta N} &=& \frac{1}{\sqrt{3}}(4\alpha -1) g_{\pi N}\;; \quad
g_{\pi\Xi} = (2\alpha -1 ) g_{\pi N}\nonumber \ ,\\
g_{\eta\Xi} &=& -\frac{1}{\sqrt{3}}(1+2\alpha ) g_{\pi N}\;;  \quad
g_{\pi\Sigma}= 2\alpha\ g_{\pi N} \nonumber\ ,\\
g_{\eta\Sigma} &=& \frac{2}{\sqrt{3}}(1-\alpha )g_{\pi N}\ .
\label{eq:su3-relation-swart}
\end{eqnarray}

To see how these relations are reflected in the OPE of
the $\gamma_5 \sigma_{\mu\nu}q^\mu p^\nu$
sum rules (see the Appendix~\ref{app:coupling-qsr} for the OPE), we 
take the SU(3) symmetric limit to organize them
in terms of two terms ${\cal O}_1$ and ${\cal O}_2$ defined as
\begin{eqnarray}
{\cal O}_1\cdot e^{-m_B^2/M^2} \quad &\equiv \quad &
\frac{1}{96\pi ^2f_\pi} (-2+4t-2t^2) \qq M^4 E_0(x) 
%
-\frac{f_\pi}{3}(-1+t^2) \qq M^2  \nonumber\\
&&-\frac{1}{54}f_\pi \delta^2 7(-1+t^2) \qq 
+\frac{1}{72\cdot 12f_\pi} (-1+2t-t^2) \qq\GG  \nonumber\\
&&+\frac{f_\pi}{72} 7(-1+t^2) m_0^2 \qq \ , \label{eq:O1} \\
{\cal O}_2\cdot e^{-m_B^2/M^2} \quad &\equiv \quad &
\frac{1}{96\pi ^2f_\pi} 12(1-t^2) \qq M^4 E_0(x) 
%
+\frac{2 f_\pi}{3}(t-t^2) \qq M^2  \nonumber\\
&&-\frac{1}{27}f_\pi \delta^2 (3-13t+10t^2) \qq 
%
+\frac{1}{48f_\pi} (1-t^2) \qq\GG  \nonumber\\
&&+\frac{f_\pi}{36} (1-3t+2t^2) m_0^2 \qq  \ . 
\label{eq:O2}
\end{eqnarray}

Specifically, we have
\begin{eqnarray}
g_{\pi N}\cdot \lambda_N^2 (1+A_{\pi N} M^2)
= &  {\cal O}_1  & + {\cal O}_2\ ,  \nonumber\\[2.5mm]
\sqrt{3}g_{\eta N}\cdot \lambda_N^2 (1+A_{\eta N} M^2)
= & -{\cal O}_1  & + {\cal O}_2\ ,  \nonumber\\[2.5mm]
g_{\pi\Xi}\cdot \lambda_N^2 (1+A_{\pi\Xi} M^2)
= & -{\cal O}_1  & \ ,              \nonumber\\[2.5mm]
\sqrt{3}g_{\eta\Xi}\cdot \lambda_N^2 (1+A_{\eta\Xi} M^2)
= & -{\cal O}_1  & -2 {\cal O}_2 \ , \nonumber\\[2.5mm]
g_{\pi\Sigma}\cdot \lambda_N^2 (1+A_{\pi\Sigma} M^2)
= &              & \ \ {\cal O}_2 \ ,  \nonumber\\[2.5mm]
\sqrt{3}g_{\eta\Sigma}\cdot \lambda_N^2 (1+A_{\eta\Sigma} M^2)
=& 2{\cal O}_1& + {\cal O}_2 \ . 
\label{eq:su3-relation-qsr}
\end{eqnarray}
Note that another SU(3) relation $\lambda_N = \lambda_\Xi = \lambda_\Sigma$
has been used in writing these equations.
Neglecting the unknown single pole term $A_{{\cal M}B}$,
we identify the $F/D$ ratio in terms of the OPE,
\begin{eqnarray}
2\alpha \sim \frac{{\cal O}_2}{{\cal O}_1 + {\cal O}_2}
\quad \longrightarrow \quad
F/D \sim \frac{{\cal O}_2}{2{\cal O}_1 + {\cal O}_2} \ .
\label{eq:F/D}
\end{eqnarray}
This is an obvious  consequence of using the
baryon currents constructed according to the SU(3) symmetry.
Hence, it provides the consistency of our sum rules with the SU(3)
relations for the couplings. 

To determine the
$F/D$ ratio, however, the unknown single pole term $A_{{\cal M}B}$
should be taken 
into account. For that purpose, we linearly fit the
RHS of Eq.~(\ref{eq:su3-relation-qsr}) and determine 
$\left[ g_{{\cal M}B}\lambda_B^2(t) \right]_{\mbox{\scriptsize fitted}}$
for a given $t$.   Once two of 
$\left[ g_{{\cal M}B}\lambda_B^2(t) \right]_{\mbox{\scriptsize fitted}}$
are determined, their ratio can be converted to yield the $F/D$ ratio
according to Eq.~(\ref {eq:su3-relation-swart}).

\begin{figure}[hbtp]
 \caption{The $F/D$ ratio from the $\gamma_5\sigma_{\mu\nu}q^\mu p^\nu$ 
   structure is plotted as a function of $\cos\theta$,
   where $\theta$ is defined as $\tan\theta = t$. 
   (See the text.)
   Corresponding $t$ is also shown at the top of the figure.
   Circles are obtained with 
   $0.65 \leq M^2 \leq 1.24\ {\rm GeV^2}$ and the continuum
   threshold $S_0 = 2.07\ {\rm GeV^2}$, triangles ;
   $0.65 \leq M^2 \leq 1.24\ {\rm GeV^2}$, $S_0 = 2.57\ {\rm GeV^2}$,
   squares ;
   $0.90 \leq M^2 \leq 1.50\ {\rm GeV^2}$, $S_0 = 2.07\ {\rm GeV^2}$.
   In the realistic region 
   $-0.78 \protect\simleq \cos\theta \protect\simleq 0.61$,
   the $F/D$ ratio is insensitive to $t$.
}
  \epsfysize=85mm
  \centerline{\epsfbox{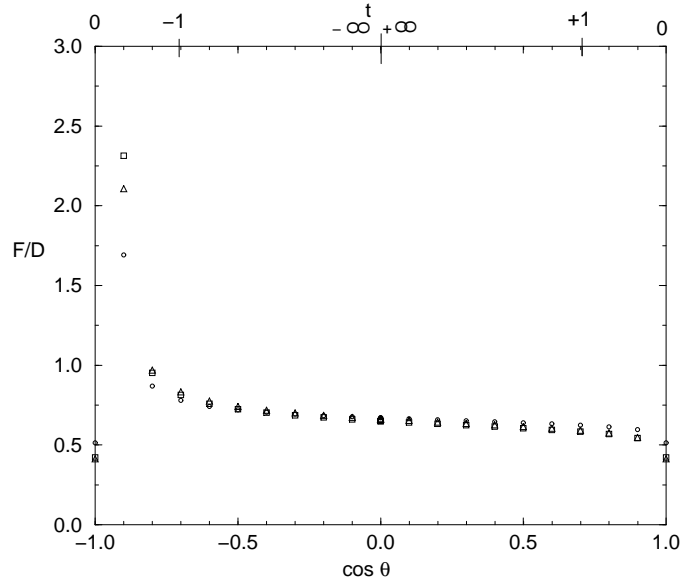}}

 \label{fig:F/D-T}
\eject
\end{figure}

In Fig.~\ref{fig:F/D-T}, the $F/D$ ratio is plotted as a function of 
$\cos\theta$. Here, to investigate the whole range of 
$ -\infty \leq t \leq +\infty $,
we introduce a new parameter $\theta$ defined as
\begin{eqnarray}
\tan\theta = t \ . 
\end{eqnarray}
Thus, the range $0 \leq t \leq +\infty $ 
corresponds to $ 0 \leq \theta \leq \pi / 2$ while
the range $ -\infty \leq t \leq 0 $
spans $ \pi / 2 \leq \theta \leq \pi $.
In Fig.~\ref{fig:F/D-T}, circles are obtained from 
the Borel window $ 0.65 \leq M^2 \leq 1.24\ {\rm GeV^2}$
with 
the continuum threshold $S_0 = 2.07\ {\rm GeV^2}$.
To see the sensitivity to this choice, we also
calculate the ratio using
(1) $ 0.65 \leq M^2 \leq 1.24\ {\rm GeV^2}$, $S_0 = 2.57\ {\rm GeV^2}$ 
(triangles),
(2) $ 0.90 \leq M^2 \leq 1.50\ {\rm GeV^2}$, $S_0 = 2.07\ {\rm GeV^2}$
(squares).
We see that the $F/D$ ratio is insensitive to 
the continuum threshold, agreeing with the discussion 
in Section~\ref{sec:s0-dep}.
Also, the calculated $F/D$ ratio 
is relatively insensitive to the choice of the Borel window. 
The peak around $ t\sim -0.5\ (\cos\theta \sim -0.9)$ can be understood 
from
Fig.~\ref{fig:t_dep-T}.  Most curves are zero around this $t$
but not simultaneously. 
The $F/D$ ratio is basically obtained
by taking a ratio of any two curves but the
ratio of the two curves around $t\sim -0.5\ (\cos\theta \sim -0.9)$ 
is not well-behaved.
On the other hand, at $t=1\ (\cos\theta = 1/\sqrt{2})$,
the $F/D$ ratio does not diverge because 
all curves for the couplings in Fig.~\ref{fig:t_dep-T}
go to zero linearly in  $(t-1)$.

The strong sensitivity of the $F/D$ ratio to $t$ within 
the region 
$ -0.5 \simleq t \simleq 1\ 
( \cos\theta \simleq -0.9, \mbox{or}\ 0.7 \simleq \cos\theta )$ 
is unrealistic because first of all, absolute total value of the OPE in
each coupling is very small in this region.
The convergence of the OPE may not be sufficient enough.
Secondly,  the
strength $\lambda_N^2$ as can be seen from 
Fig.~\ref{fig:t_dep-mass-qsr} is negative, thus not physical.
Therefore, a reasonable value for the $F/D$ ratio
should be obtained away from this region.
We moderately take the realistic region as 
(1) $ t \simleq -0.8\ ( -0.78 \simleq \cos\theta )$ 
and (2) $1.3 \simleq t\ ( \cos\theta \simleq 0.61 )$.
The former constraint gives us the maximum value of
$F/D \sim 0.84$, and the latter constraint gives us
the minimum value of $F/D \sim 0.63$.
Therefore, we conclude $F/D \sim 0.6 - 0.8$.
This range includes the value from the SU(6) quark
model ($F/D = 2/3$),
and is slightly higher than that extracted from 
semi-leptonic decay rates of 
hyperons ($F/D \sim 0.57$)~\cite{ratcliffe}.
It is often argued that the choice of $t=-1$ (the Ioffe current)
is optimal because the instanton effect~\cite{griegel} and 
the continuum contribution~\cite{leinweber} is small, 
and the chiral breaking effects are maximized.
If we choose $t\sim -1$, our estimate becomes 
$F/D \sim 0.76 - 0.81 $, that is somewhat larger than 
the SU(6) value.

\begin{figure}[hbtp]
 \caption{The $F/D$ ratio from the $i\gamma_5$ structure 
   is plotted as a function of $\cos\theta$. See the caption of 
   Fig.~\ref{fig:F/D-T} for the explanation of each symbol. 
   The $F/D$ ratio is sensitive to $t$ and  we can not predict 
   reliable value.
}
  \epsfysize=85mm
  \centerline{\epsfbox{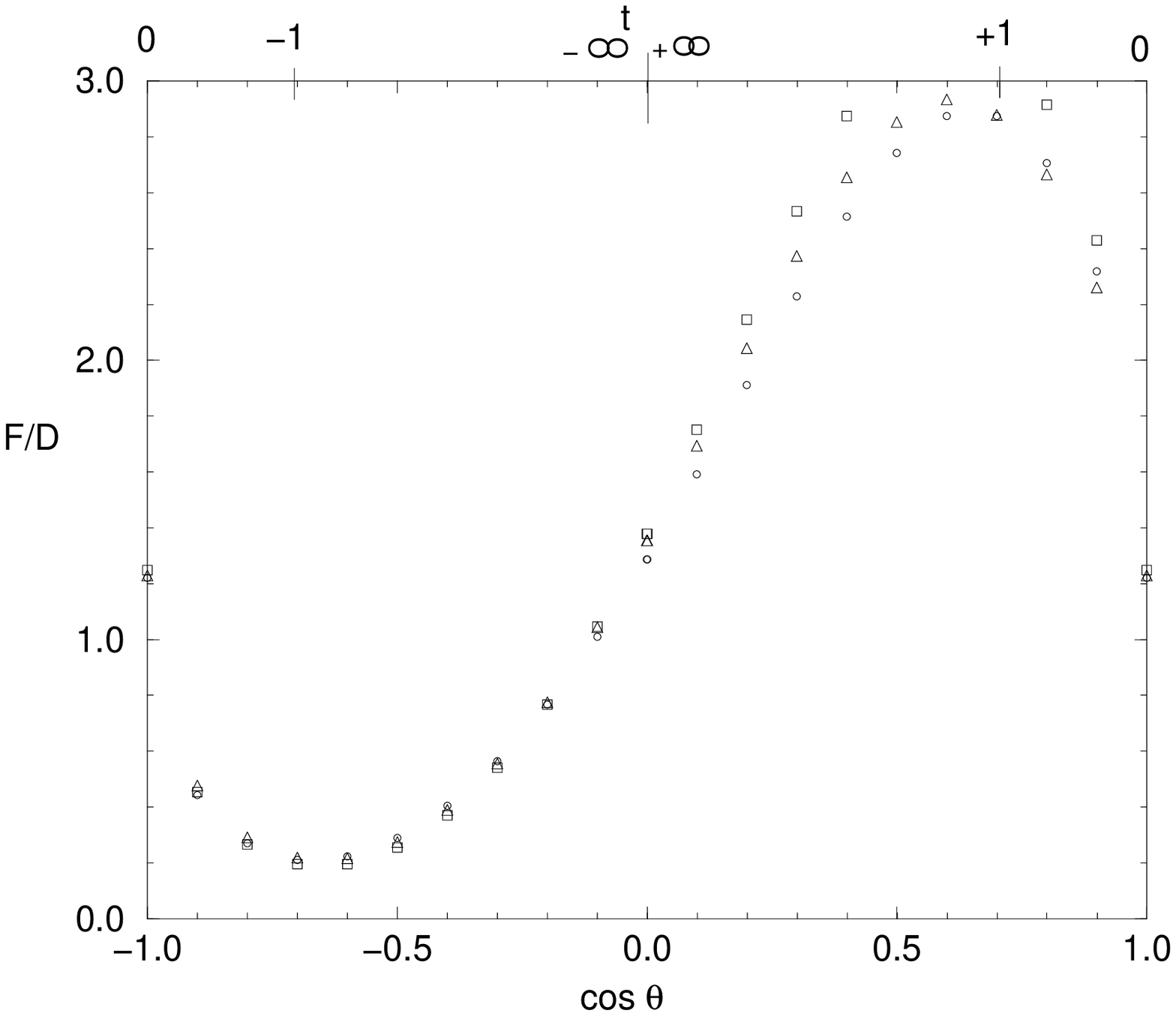}}

 \label{fig:F/D-PS}
\end{figure}

As a comparison, let us briefly consider the $i\gamma_5$ structure case.
In this case too, we can classify the OPE of the
Appendix~\ref{subapp:PS-qsr} according to 
Eq.~(\ref{eq:su3-relation-swart}) and identify the terms responsible
for the $F/D$ ratio.
By taking similar steps as T sum rules,
we determine the $F/D$ ratio.
Fig.~\ref{fig:F/D-PS} shows the $F/D$ ratio as a function of
$\cos\theta$. Compared with Fig.~\ref{fig:F/D-T},
the $F/D$ ratio is very sensitive to $t$.
As discussed in Section~\ref{sec:current-dep}, 
$f^{\rm OPE}_{\rm rest}$ may cause this huge $t$-dependence.
Another possibility is due to the large contribution from direct 
instanton in the pseudoscalar channel.
The direct instanton effect is believed to cause large OZI 
breaking in $\eta$ and $\eta '$.
To confirm it, it will be necessary to include the direct instanton 
effect in this pseudoscalar channel.
Nevertheless, the correlation function Eq.~(\ref{tmcor}) is often used
in literature to calculate various couplings and
our study suggests that one has to be careful in choosing
a Dirac structure in that correlation function.



\section{Conclusions}
\label{sec:summary}

In this work, we calculated the correlation function Eq.~(\ref{tmcor})
for the vertices, $\pi NN$, $\eta NN$, $\pi\Xi\Xi$, $\eta\Xi\Xi$, 
$\pi\Sigma\Sigma$, and $\eta\Sigma\Sigma$, using QCD sum rules. 
In the construction of sum rules, we used
general baryon currents with no derivative instead of the Ioffe current,
which enables us to discuss the dependence of sum rule results on currents.
We proposed a new criterion to choose a pertinent
Dirac structure by studying the dependence of the 
correlation function on the baryon currents.
Specifically, it is imposed that a physical parameter is 
ideally independent of a chosen current.  
In checking this constraint, the SU(3) symmetric limit
is quite useful as it provides simple relations among 
the couplings.  
It is found that the $\gamma_5\sigma_{\mu\nu}q^\mu p^\nu$ structure
satisfies the ideal constraint relatively well, which
moderately restricts the $F/D$ ratio within the range, $F/D\sim 0.6-0.8$.
However, the $i\gamma_5$ sum rules beyond the
chiral limit do not satisfy the constraint, which
provides a large window for the value
of the $F/D$ ratio depending on currents.

In the present study, we considered only the SU(3) limit
of the meson-baryon couplings. In fact, the OPE for the
$\gamma_5\sigma_{\mu\nu}q^\mu p^\nu$ structure given 
in the Appendix~\ref{subapp:T-qsr} contain effects of SU(3) breaking
partially as 
$m_N \neq m_\Xi \neq m_\Sigma$, 
$\lambda_N \neq \lambda_\Xi \neq \lambda_\Sigma$,
$\qq \neq \ss$ and
$f_\pi \neq f_\eta$.
If we include these differences, obtained coupling constants
break the SU(3) symmetry accordingly.
We, however, do not quantify this because other sources of SU(3)
breaking are expected.
Especially, the large strange quark mass $(m_s)$ may cause 
non-negligible SU(3) breaking effects.
So far, the OPE for the $\gamma_5\sigma_{\mu\nu}q^\mu p^\nu$
structure is truncated to ${\cal O}(p)$ so that it is consistent
with the chiral expansion, while effects of $m_s$ can only be included 
at ${\cal O}(p^2)$.
In order to quantify SU(3) breaking effects on the meson-baryon couplings,
it will be necessary to include ${\cal O}(p^2)$ contribution.
The present formulation may give a solid starting point for 
such analyses in future.


\acknowledgments

This work was supported in part by
the Grant-in-Aid for scientific
research (C) (2) 11640261 of  the Ministry of Education, 
Science, Sports and Culture of Japan.
H. Kim was supported by the Brain Korea 21 project.
We would like to thank Prof. S.H. Lee for useful discussions.


\begin{appendix}

\section{Coupling sum rules from the PS, T, PV structure}
\label{app:coupling-qsr}

Coupling sum rules for 
$\pi NN$, $\eta NN$, $\pi\Xi\Xi$, $\eta\Xi\Xi$, 
$\pi\Sigma\Sigma$, and $\eta\Sigma\Sigma$
are presented here. 
For the $\eta$ couplings,
$\eta -\eta '$ mixing is not introduced because our analysis in this
paper is within SU(3).
In the OPE side, the quark-gluon mixed condensate is parametrized as
$\braket{\bar{q_i}g_s\sigma\cdot {\cal G}q_i} 
\equiv m_0^2 \braket{\bar{q_i}q_i}$ where $q_i = u,d,s$-quark.
Also, we take the isospin symmetric limit,
$\braket{\bar{u}u} = \braket{\bar{d}d} \equiv \qq$
and $ m_u = m_d \equiv m_q$.
The continuum contribution is denoted by the factor,
$E_n(x\equiv S_0/M^2) = 1-(1+x+\cdots +x^n/n!)e^{-x}$
where $S_0$ is the continuum threshold.

\subsection{Coupling sum rules from the $i\gamma_5$ structure}
\label{subapp:PS-qsr}

Here we present the $i\gamma_5$ sum rules  up to
dimension 8 constructed at the order $p^2 = m_\pi^2$.
$A_{{\cal M}B}^{\mbox{\tiny PS}}$ denotes the 
unknown single-pole term coming from transitions between the
ground state baryon and higher resonance states.

%
%
\begin{eqnarray}
%
%
%
%
\lefteqn{
g_{\pi N}m_\pi ^2\lambda _N^2(1+A^{\mbox{\tiny PS}}_{\pi N} M^2) 
e^{-m_N^2/M^2}
} \nonumber\\
&=&
-\frac{m_\pi^2}{48\pi^2f_\pi}(-5-2t+7t^2)\qq M^4 E_0(x)
+ \frac{3f_{3\pi}m_\pi^2}{16\sqrt{2}\pi^2}(-1+2t-t^2)M^4 E_0(x)   \nonumber\\
&& -\frac{1}{2f_\pi}(-5-2t-t^2)m_q \qq^2 M^2 
-\frac{m_\pi^2}{288f_\pi}(-7+2t+5t^2)\qq\GG   \nonumber\\
&& -\frac{1}{12f_\pi}(-7-6t-7t^2)m_q m_0^2 \qq ^2  \nonumber\\[3mm]
%
%
%
%
\lefteqn{
\sqrt{3}g_{\eta N}m_\eta ^2\lambda _N^2(1+A^{\mbox{\tiny PS}}_{\eta N} M^2) 
e^{-m_N^2/M^2} 
} \nonumber\\
&=&
-\frac{m_\eta^2}{48\pi^2f_\eta}(-7+2t+5t^2)\qq M^4 E_0(x)
+ \frac{3f_{3\eta}m_\eta^2}{16\sqrt{2}\pi^2}(1-2t+t^2)M^4 E_0(x) \nonumber\\
&& -\frac{1}{2f_\eta}(-7-14t-3t^2)m_q \qq^2 M^2 
-\frac{m_\eta^2}{288f_\eta}(-5-2t+7t^2)\qq\GG   \nonumber\\
&& -\frac{1}{12f_\eta}(-5-2t-5t^2)m_q m_0^2 \qq ^2  \nonumber\\[3mm]
%
%
%
%
\lefteqn{
g_{\pi \Xi}m_\pi ^2\lambda _\Xi^2(1+A^{\mbox{\tiny PS}}_{\pi \Xi} M^2) 
e^{-m_\Xi^2/M^2}
}\nonumber\\
&=&
-\frac{m_\pi^2}{48\pi^2f_\pi}(-1+2t-t^2)\qq M^4 E_0(x)
+ \frac{3f_{3\pi}m_\pi^2}{16\sqrt{2}\pi^2}(1-2t+t^2)M^4 E_0(x)\nonumber\\
&& -\frac{1}{2f_\pi}(-1-6t-t^2)m_s \qq\ss M^2 
-\frac{m_\pi^2}{288f_\pi}(1-2t+t^2)\qq\GG \nonumber\\
&& -\frac{1}{12f_\pi}(1+2t+t^2)m_s m_0^2 \qq\ss \nonumber\\[3mm]
%
%
%
%
\lefteqn{
\sqrt{3}g_{\eta \Xi}m_\eta ^2\lambda _\Xi^2
(1+A^{\mbox{\tiny PS}}_{\eta \Xi} M^2) e^{-m_\Xi^2/M^2}
}\nonumber\\
&=&
-\frac{m_\eta^2}{48\pi^2f_\eta}
[\ (-1+2t-t^2)\qq + 12(1-t^2)\ss\ ] M^4 E_0(x)
+ \frac{3f_{3\eta}m_\eta^2}{16\sqrt{2}\pi^2}(1-2t+t^2)M^4 E_0(x)\nonumber\\
&& -\frac{1}{2f_\eta}
        [\ -(1+6t+t^2)m_s\qq +2(3+4t+t^2)(m_q\qq+m_s\ss )\ ] \ \ss M^2 
\nonumber\\
&& -\frac{m_\eta^2}{288f_\eta}
[\ (1-2t+t^2)\qq + 12(1-t^2)\ss\ ] \GG \nonumber\\
&& -\frac{1}{12f_\eta}
[\ (1+2t+t^2)m_s\qq + 2(3+2t+3t^2)(m_q\ss + m_s\qq )\ ] \ m_0^2 \ss 
\nonumber\\[3mm]
%
%
%
%
\lefteqn{
g_{\pi\Sigma}m_\pi ^2\lambda _\Sigma^2
(1+A^{\mbox{\tiny PS}}_{\pi \Sigma} M^2) 
e^{-m_\Sigma^2/M^2}
}\nonumber\\
&=&
-\frac{m_\pi^2}{48\pi^2f_\pi}(-6+6t^2)\qq M^4 E_0(x)
-\frac{1}{2f_\pi}(-3-4t-t^2) (m_q\qq + m_s\ss ) \qq M^2 \nonumber\\
&& -\frac{m_\pi^2}{288f_\pi}(-6+6t^2)\qq\GG 
-\frac{1}{12f_\pi}(-3-2t-3t^2) (m_q\ss + m_s\qq ) m_0^2 \qq \nonumber\\[3mm]
%
%
%
%
\lefteqn{
\sqrt{3}g_{\eta\Sigma}m_\eta ^2
\lambda _\Sigma^2(1+A^{\mbox{\tiny PS}}_{\eta\Sigma} M^2) e^{-m_\Sigma^2/M^2}
}\nonumber\\
&=&
-\frac{m_\eta^2}{48\pi^2f_\eta}
[\ -6(1-t^2)\qq + 2(1-2t+t^2)\ss\ ] M^4 E_0(x) 
+ \frac{3f_{3\eta}m_\eta^2}{16\sqrt{2}\pi^2}
(-2+4t-2t^2)M^4 E_0(x) \nonumber\\
&& -\frac{1}{2f_\eta}
[\ 2(1+6t+t^2)m_q\ss - (3+4t+t^2)(m_q\qq+m_s\ss )\ ]\ \qq M^2 \nonumber\\
&& -\frac{m_\eta^2}{288f_\eta}
[\ -6(1-t^2)\qq + 2(-1+2t-t^2)\ss\ ] \GG \nonumber\\
&& -\frac{1}{12f_\eta}
[\ -2(1+2t+t^2)m_q\ss -(3+2t+3t^2)(m_q\ss + m_s\qq )\ ] m_0^2 \qq 
\label{eq:PS-qsr}
\end{eqnarray}
%
%

\subsection{Coupling sum rules from the $\gamma_5\sigma_{\mu\nu}q^\mu p^\nu$ 
structure}
\label{subapp:T-qsr}

The $\gamma_5\sigma_{\mu\nu}q^\mu p^\nu$ sum rules
up to dimension 7  are the following.  
Again, $A^{\mbox{\tiny T}}_{{\cal M}B}$ denotes the unknown single-pole term
contribution.
%
%
\begin{eqnarray}
%
%
%
%
\lefteqn{
g_{\pi N}\lambda _N^2 (1+ A^{\mbox{\tiny T}}_{\pi N}M^2) e^{-m_N^2/M^2}
} \nonumber\\
&=&
\frac{1}{96\pi ^2f_\pi} (10+4t-14t^2) \qq M^4 E_0(x) 
-\frac{f_\pi}{3}(-1-2t+3t^2) \qq M^2  \nonumber\\
&&-\frac{1}{54}f_\pi \delta^2 (-1-26t+27t^2) \qq 
+\frac{1}{72\cdot 12f_\pi} (17+2t-19t^2) \qq\GG  \nonumber\\
&&+\frac{f_\pi}{72} (-5-6t+11t^2) m_0^2 \qq  \nonumber\\[3mm]
%
%
%
%
\lefteqn{
\sqrt{3}g_{\eta N}\lambda _N^2 (1+ A^{\mbox{\tiny T}}_{\eta N}M^2) 
e^{-m_N^2/M^2}
} \nonumber\\
&=&
\frac{1}{96\pi ^2f_\eta} (14-4t-10t^2) \qq M^4 E_0(x) 
-\frac{f_\eta}{3}(1-2t+t^2) \qq M^2  \nonumber\\
&&-\frac{1}{54}f_\eta \delta^2 (13-26t+13t^2) \qq 
+\frac{1}{72\cdot 12f_\eta} (19-2t-17t^2) \qq\GG  \nonumber\\
&&+\frac{f_\eta}{72} (9-6t-3t^2) m_0^2 \qq \nonumber\\[3mm]
%
%
%
%
\lefteqn{
g_{\pi \Xi}\lambda _\Xi^2 (1+ A^{\mbox{\tiny T}}_{\pi \Xi}M^2) 
e^{-m_\Xi^2/M^2}
} \nonumber\\
&=&
\frac{1}{96\pi ^2f_\pi} (2-4t+2t^2) \qq M^4 E_0(x) 
-\frac{f_\pi}{3}(1-t^2) \ss M^2  \nonumber\\
&&-\frac{1}{54}f_\pi \delta^2 (7-7t^2) \ss 
+\frac{1}{72\cdot 12f_\pi} (1-2t+t^2) \qq\GG  \nonumber\\
&&+\frac{f_\pi}{72} (7-7t^2) m_0^2 \ss \nonumber\\[3mm]
%
%
%
%
\lefteqn{
\sqrt{3}g_{\eta \Xi}\lambda _\Xi^2 (1+ A^{\mbox{\tiny T}}_{\eta \Xi}M^2) 
e^{-m_\Xi^2/M^2}
}\nonumber\\
&=&
\frac{1}{96\pi ^2f_\eta} 
\left[\  (2-4t+2t^2) \qq + 24(-1+t^2)\ss \ \right] M^4 E_0(x) \nonumber\\
&& -\frac{f_\eta}{3} 
\left[\  (-2+4t-2t^2) \qq + 3(1-t^2) \ss \ \right] M^2  \nonumber\\
&& -\frac{1}{54}f_\eta \delta^2 
\left[\  26(-1+2t-t^2)\qq + 21(1-t^2) \ss \ \right]  \nonumber\\
&& +\frac{1}{72\cdot 12f_\eta} 
\left[\  (1-2t+t^2) \qq + 36(-1+t^2) \ss \ \right] \GG  \nonumber\\
&& +\frac{f_\eta}{72} 
\left[\  6(-1+2t-t^2)\qq + 9(1-t^2)\ss \ \right]m_0^2 \nonumber\\[3mm]
%
%
%
%
\lefteqn{
g_{\pi \Sigma}\lambda _\Sigma^2 (1+ A^{\mbox{\tiny T}}_{\pi \Sigma}M^2) 
e^{-m_\Sigma^2/M^2}
}\nonumber\\
&=&
\frac{1}{96\pi ^2f_\pi} 12(1-t^2) \qq  M^4 E_0(x) 
-\frac{f_\pi}{3}
\left[\ (-1+t^2)\qq + (1-2t+t^2)\ss \ \right] M^2  \nonumber\\
&&-\frac{1}{54}f_\pi \delta^2 
\left[\ 7(-1+t^2)\qq + 13(1-2t+t^2)\ss \ \right] 
+\frac{1}{72\cdot 12f_\pi} 18(1-t^2) \qq\GG  \nonumber\\
&&+\frac{f_\pi}{72} 
\left[\ (-1+t^2)\qq + 3(1-2t+t^2)\ss \ \right] m_0^2 \nonumber\\[3mm]
%
%
%
%
\lefteqn{
\sqrt{3}g_{\eta \Sigma}\lambda _\Sigma^2 
(1+ A^{\mbox{\tiny T}}_{\eta \Sigma}M^2) e^{-m_\Sigma^2/M^2}
}\nonumber\\
&=&
\frac{1}{96\pi ^2f_\eta} 
\left[\  12(1-t^2) \qq + 4(-1+2t-t^2)\ss \ \right] M^4 E_0(x) \nonumber\\
&& -\frac{f_\eta}{3} 
\left[\  3(-1+t^2) \qq + (1-2t+t^2) \ss \ \right] M^2  \nonumber\\
&&-\frac{1}{54}f_\eta \delta^2 
\left[\  21(-1+t^2)\qq + 13(1-2t+t^2) \ss \ \right] \nonumber\\
&& +\frac{1}{72\cdot 12f_\eta} 
\left[\  18(1-t^2) \qq + 2(-1+2t-t^2) \ss \ \right] \GG  \nonumber\\
&&+\frac{f_\eta}{72} 
\left[\  15(-1+t^2)\qq + 3(1-2t+t^2)\ss \ \right]m_0^2 
\label{eq:T-qsr}
\end{eqnarray}
%
%

\subsection{Coupling sum rules from 
the $i\gamma_5\protect\fslash{p}$ structure}
\label{subapp:PV-qsr}

The $i\gamma_5\fslash{p}$ sum rules  
up to dimension 7 
are presented here.
%
%
\begin{eqnarray}
%
%
%
%
\lefteqn{
g_{\pi N}m_N \lambda _N^2(1+A^{\mbox{\tiny PV}}_{\pi N} M^2) e^{-m_N^2/M^2}
}\nonumber\\
&=&
\frac{f_\pi}{24\pi ^2}(10+8t+10t^2)M^6 E_1(x) 
-\frac{f_\pi \delta ^2}{48\pi ^2}(-20-16t-20t^2)M^4 E_0(x)\nonumber\\
&& +\frac{f_\pi}{72}\GG (5+4t+5t^2) M^2 
+\frac{1}{18f_\pi} \qq ^2 (-1-2t+3t^2) M^2 \nonumber\\
&& +\frac{f_\pi \delta ^2}{36\cdot 18}\GG (-4-2t-4t^2)
+\frac{1}{432f_\pi} m_0^2 \qq ^2 (-5-6t+11t^2) \nonumber\\[3mm]
%
%
%
%
\lefteqn{
\sqrt{3}g_{\eta N}m_N \lambda _N^2(1+A^{\mbox{\tiny PV}}_{\eta N} M^2) 
e^{-m_N^2/M^2}
}\nonumber\\
&=&
\frac{f_\eta}{24\pi ^2}(8+8t+8t^2)M^6 E_1(x) 
-\frac{f_\eta \delta ^2}{48\pi ^2}(-22-4t-22t^2)M^4 E_0(x)\nonumber\\
&& +\frac{f_\eta}{72}\GG (5+2t+5t^2) M^2 
+\frac{1}{18f_\eta} \qq ^2 (1-2t+t^2) M^2 \nonumber\\
&& +\frac{f_\eta \delta ^2}{36\cdot 18}\GG (-10-4t-10t^2)
+\frac{1}{432f_\eta} m_0^2 \qq ^2 (9-6t-3t^2) \nonumber\\[3mm]
%
%
%
%
\lefteqn{
g_{\pi\Xi}m_\Xi \lambda _\Xi^2(1+A^{\mbox{\tiny PV}}_{\pi\Xi} M^2) 
e^{-m_\Xi^2/M^2}
}\nonumber\\
&=&
\frac{f_\pi}{24\pi ^2}(-1-t^2)M^6 E_1(x) 
-\frac{f_\pi \delta ^2}{48\pi ^2}(-1+6t-t^2)M^4 E_0(x)\nonumber\\
&& +\frac{f_\pi}{72}\GG (-t) M^2 
+\frac{1}{18f_\pi} \qq\ss (1-t^2) M^2 \nonumber\\
&& +\frac{f_\pi \delta ^2}{36\cdot 18}\GG (-3-t-3t^2)
+\frac{1}{432f_\pi} m_0^2 \qq\ss (7-7t^2) \nonumber\\[3mm]
%
%
%
%
\lefteqn{
\sqrt{3}g_{\eta\Xi}m_\Xi \lambda _\Xi^2(1+A^{\mbox{\tiny PV}}_{\eta\Xi} M^2) 
e^{-m_\Xi^2/M^2}
}\nonumber\\
&=&
\frac{f_\eta}{24\pi ^2}(-19-16t-19t^2)M^6 E_1(x) 
-\frac{f_\eta \delta ^2}{48\pi ^2}(41+26t+41t^2)M^4 E_0(x)\nonumber\\
&& +\frac{f_\eta}{72}\GG (-10-7t-10t^2) M^2 
+\frac{1}{18f_\eta} 
\left[\ (3-3t^2)\qq + (-2+4t-2t^2)\ss \ \right] \ss M^2 \nonumber\\
&& +\frac{f_\eta \delta ^2}{36\cdot 18}\GG (11+5t+11t^2)
+\frac{1}{432f_\eta} 
\left[\ 9(1-t^2)\qq + 6(-1+2t-t^2)\ss\ \right] m_0^2 \ss \nonumber\\[3mm]
%
%
%
%
\lefteqn{
g_{\pi\Sigma}m_\Sigma \lambda _\Sigma^2
(1+A^{\mbox{\tiny PV}}_{\pi\Sigma} M^2) 
e^{-m_\Sigma^2/M^2}
}\nonumber\\
&=&
\frac{f_\pi}{24\pi ^2}(9+8t+9t^2)M^6 E_1(x) 
-\frac{f_\pi \delta ^2}{48\pi ^2}(-21-10t-21t^2)M^4 E_0(x)\nonumber\\
&& +\frac{f_\pi}{72}\GG (5+3t+5t^2) M^2 
+\frac{1}{18f_\pi} 
\left[\ (1-2t+t^2)\qq + (-1+t^2)\ss \ \right]\qq M^2 \nonumber\\
&& +\frac{f_\pi \delta ^2}{36\cdot 18}\GG (-7-3t-7t^2)
+\frac{1}{432f_\pi} 
\left[\ -3(-1+2t-t^2)\qq - (1-t^2)\ss \ \right] m_0^2 \qq \nonumber\\[3mm]
%
%
%
%
\lefteqn{
\sqrt{3}g_{\eta\Sigma}m_\Sigma \lambda _\Sigma^2
(1+A^{\mbox{\tiny PV}}_{\eta\Sigma} M^2) 
e^{-m_\Sigma^2/M^2}
}\nonumber\\
&=&
\frac{f_\eta}{24\pi ^2}(11+8t+11t^2)M^6 E_1(x) 
-\frac{f_\eta \delta ^2}{48\pi ^2}(-19-22t-19t^2)M^4 E_0(x)\nonumber\\
&& +\frac{f_\eta}{72}\GG (5+5t+5t^2) M^2 
+\frac{1}{18f_\eta} 
\left[\ (1-2t+t^2)\qq -3(1-t^2)\ss \ \right]\qq M^2 \nonumber\\
&& +\frac{f_\eta \delta ^2}{36\cdot 18}\GG (-1-t-t^2)
+\frac{1}{432f_\eta} 
\left[\ -3(-1+2t-t^2)\qq - 15(1-t^2)\ss \ \right] m_0^2 \qq
\label{eq:PV-qsr}
\end{eqnarray}
%
%

\end{appendix}




\begin{references}
\bibitem{SVZ}	    {M.A. Shifman, A.I. Vainshtein, and V.I. Zakharov,
                            Nucl. Phys. {\bf B 147} (1979) 385, 448.}
\bibitem{SH}       {H. Shiomi and T. Hatsuda,
                            Nucl. Phys. {\bf A 594} (1995) 294.}
\bibitem{BK}    {M. C. Birse and B. Krippa,
                            Phys. Lett. B {\bf 373} (1996) 9;
                            Phys. Rev. C {\bf 54} (1996) 3240.}
\bibitem{hung1}     {H. Kim, S. H. Lee and M. Oka,
                    Phys. Lett. B{\bf 453} (1999) 199.}
\bibitem{hung2}     {H. Kim, S. H. Lee and M. Oka,
                    Phys. Rev. D{\bf 60} (1999) 034007.}
\bibitem{hung5}     {H. Kim, 
                     Eur. Phys. Jour. A7,(2000), 121.
                     }
\bibitem{KD1}     {H. Kim, T. Doi, M. Oka and S. H. Lee,
                            Nucl. Phys.  {\bf A662} (2000) 371.}
\bibitem{KD2}     {H. Kim, T. Doi, M. Oka and S. H. Lee,
			Nucl. Phys.  {\bf A678} (2000) 295.;nucl-th/0002011.}
\bibitem{ioffe-smilga} {B. L. Ioffe and A. V. Smilga,
                            Nucl. Phys.  {\bf B232} (1984) 109.}
\bibitem{swart}     {J. J. de Swart, 
                    Rev. Mod. Phys. {\bf 35} (1963) 916.; {\bf 37} 
                    (1965) 326 (E).}
\bibitem{rijken1}	{V. G. J. Stoks and Th. A. Rijken,
				Phys. Rev. C {\bf 59} (1999) 3009.}
\bibitem{rijken2}   {Th. A. Rijken, V. G. J. Stoks and Y. Yamamoto,
                    Phys. Rev. C {\bf 59} (1999) 21.}
\bibitem{bely-kogan}   {V. M. Belyaev and Ya. I. Kogan,
                   JETP Lett. {\bf 37} (1983) 730;
                   B. L. Ioffe and A. G. Oganesian,
                   Phys. Rev. D {\bf 57} (1998) R6590.}
\bibitem{nishikawa-kondo} {T. Nishikawa, S. Saito and Y. Kondo,
			   Phys. Rev. Lett. {\bf 84} (2000) 2326.}
\bibitem{ioffe}     {B. L. Ioffe,
                            Nucl. Phys.  {\bf B188} (1981) 317.}
\bibitem{qsr}       { L.J. Reinders, H. Rubinstein and S. Yazaki, 
	                   Phys. Rep. {\bf 127} (1985) 1.}
\bibitem{com}       {H. Kim, Phys. Rev. C {\bf 61} (2000) 019801.}
\bibitem{single}    {H. Kim, nucl-th/9906081, Prog. Theo. Phys. {\bf 103}
                    (2000) 1001.}
\bibitem{leinweber}   {D.B. Leinweber,
                            Phys. Rev. D {\bf 51} (1995) 6383.}
\bibitem{ratcliffe}     {P. G. Ratcliffe,
                    Phys. Lett. B{\bf 365} (1996) 383.; 
                    hep-ph/9710458.}
\bibitem{griegel}   {D. K. Griegel, Ph.D thesis,
                     University of Maryland (1991).}

\end{references}
\end{document}